\documentstyle[epsf,12pt]{article}
\newlength{\dinwidth}
\newlength{\dinmargin}
\input epsf
\newcommand{\vektor}[1]{\mbox{\boldmath $#1$}}
\newcommand{\NM}[1]{\mbox{$\cal #1$}}

\begin{document}
\vspace{2 cm}
\title{
{\bf A novel approach to error function
minimization for feedforward neural networks
 }\\
\author{Ralph Sinkus\thanks{I. Institut f\"ur Experimentalphysik der
Universit\"at Hamburg, Germany. This work has been supported by the
German Federal Minister for Research and Technology (BMFT), contract Nr.
6HH19I}\\
}}

\maketitle
\vspace{5 cm}
\begin{abstract}
Feedforward neural networks with error backpropagation (FFBP) are widely
applied to pattern recognition. One general problem encountered with
this type of neural
networks is the uncertainty, whether the minimization procedure has converged
to a
global minimum of the cost function.
To overcome this problem a novel approach to minimize the error
function is presented.
It allows to monitor the approach to the global minimum and as an outcome
several ambiguities related to the choice of free parameters
of the minimization procedure are removed.
\end{abstract}
\vspace{-20cm}
\begin{flushleft}
\tt DESY 94-182 \hfill ISSN 0418-9833\\
October 1994\\
\end{flushleft}
\setcounter{page}{0}
\thispagestyle{empty}
\newpage
\section{Introduction}
In high energy physics the separation of signal to background usually
turns out to be a multi-dimensional classification problem with many variables
involved in order to achieve a reasonable rejection factor. This is the
domain, where neural networks with their intrinsic ability to deal with
many dimensions, are worth to be applied.
The output values of neural networks (NN) can be interpreted as estimators
of {\it a posteriori} Bayesian probabilities which provide the link
to classification problems of higher order \cite{baypro}.\\

A neural network can be regarded as a non-linear combination of several
transformation matrices, with entries (denoted as weights) adjusted in the
training phase by a least squares minimization of an error function.
There are several technical problems associated with the training of
the neural network. Real world applications rarely allow a
perfect separation of patterns, i.e. given a problem with patterns of
two classes $C_{1}$ and $C_{2}$, a certain fraction of patterns belonging
to class $C_{1}$ will look like patterns from class $C_{2}$ and vice versa.
This effect may be denoted as the {\it confusing teacher problem}. In high
energy physics one deals with overlapping distributions which will cause an
{\it a priori} contamination, i.e. indistinguishable patterns assigned to
different classes. The non-linearity of the problem
and the number of free parameters involved, enhance the possibility of
the minimization procedure to converge to a local minimum of the error
function.
This leads to a deterioration of the separation ability and therefore a poorer
estimate of the Bayesian probabilities, which are lower limits to the
probability
of error in any problem of classification.
The type of network used in this analysis is known as feedforward neural
network
with error backpropagation  (FFBP) \cite{first_FFNN_backpro}.
The name originates from the specific architecture
of the transformation matrices and the method applied to optimize their
entries.
A pattern is represented by an input vector whose entries are
processed by the network in several layers of units. Each unit feeds only
the units of the following layer (feedforward). During the minimization
procedure
the calculated difference between the actual network output and the desired
output
is used to adjust the weights (error backpropagation)
\cite{cal_trigger_application}.\\

A general description rule to avoid problems with local minima in
the minimization procedure for feedforward neural networks with a quadratic
cost function will be presented in this article.
The basic features of the new model are demonstrated by a one-dimensional
problem.

\section{Mathematical foundation of feedforward neural networks}
Each pattern to be classified is represented by a vector \vektor{X} of
dimension $K$ (called the input vector). For the purpose of estimating
the Bayesian a posteriori probabilities the input vector \vektor{X} is
projected
into the output space by means of the neural network formula
\begin{equation}
\vektor{Y}~=~f(\vektor{X})~~~~,
\end{equation}
in which \vektor{Y} is of dimension $I$, equal to the total number of possible
classes $C_{l}$, with $l \:\epsilon\: [1,I]$. Standard FFBP use for $f$
\begin{eqnarray}
f(\vektor{X}) & = &
g\left[~\frac{A_{ij}~g\left(~\frac{B_{jk}X_{k}-\alpha_{j}}{t}~\right)-\b
eta_{i}}{t}~\right] \label{nn_formula} \\
g(...)        & = & \frac{1}{2}\left(~1+\tanh(...)~\right)~~~~, \nonumber
\end{eqnarray}
with $i \:\epsilon\: [1,I]~,~j \:\epsilon\: [1,J]~,~k \:\epsilon\:
[1,K]~$ and where
the Einstein convention for same indices was used. For two overlapping Gaussian
distributions the function $g$ represents the exact solution for the Bayesian
probabilities which is the motivation
for the choice of this particular \mbox{\it sigmoid} function
\cite{approx_any_func}.
The weights $A_{ij},~B_{jk},~\alpha_{j}~{\rm and}~ \beta_{i}$ are
the free parameters of the fit function $f$, and $J$ denotes the number of
hidden nodes. In analogy to spin Ising models $t$ is called {\it temperature}
and is
usually set to one. For any classification problem the aim is to achieve
for an input vector \vektor{X} belonging to class $C_{l}$
the output value \vektor{Y} $=$ \vektor{O}$(C_{l})$, where
\vektor{O}$(C_{l})$ is a unit vector whose components are all zero
except for the entry with index $l$.
For any $A_{ij},~B_{jk},~\alpha_{j}~{\rm and}~ \beta_{i}$, initialized
at random in an
interval $[-\epsilon,+\epsilon]$, one defines the error function
\begin{equation}
E~=~\frac{1}{2\,N}~\sum\limits_{n=1}^{N}~\sum\limits_{i=1}^{I}~
\left(~Y_{i}(\vektor{X}(C_{l},n))~-~O_{i}(C_{l})~\right)^{2}~~~,
\label{error_function}
\end{equation}
where the first sum runs over all $N$ patterns available and the second
sum runs over
all possible classes $I$.
This error function is minimized iteratively by a gradient
descent method
\begin{eqnarray}
A_{ij}^{(\mu+1)}      & = & A_{ij}^{(\mu)}~-~\eta~\Delta A_{ij}^{(\mu)} \\
\Delta A_{ij}^{(\mu)} & = & \frac{\partial E^{(\mu)}}{\partial A_{ij}}~+~
                            \underbrace{\kappa~\Delta
A_{ij}^{(\mu-1)}}_{\em momentum~term}~~~~,
\end{eqnarray}
where $\eta$ is the step size $(\,\eta\:\epsilon\:(0,\infty)\,)$ and
$0\:\leq\:\kappa\:<\:1$ is the weight of the momentum term. The same procedure
applies to the matrices $B,~\alpha$ and $\beta$.
The upper index $(\mu)$ denotes the iteration step.
The momentum term serves to damp possible oscillations and to speed up
the convergence
in regions where $E$ is almost flat \cite{intro_neural_comp}.\\

Typically FFBP consist of many free parameters which need to be adjusted
by the minimization procedure. Due to the non-linearity of the
neural net function $f(\vektor{X})$ it is very likely that local minima
of the error function $E$ will prevent the convergence
to the global minimum which results in a deterioration of the classification
ability. One common method to avoid this problem is to change the
definition of $E$, i.e.
instead of averaging over all patterns $N$, the sum for $n$ in equation
(\ref{error_function}) extends only over
$\tilde N$ patterns with $1\:\leq\:\tilde N\:<\:N$ (called
\mbox{\it incremental updating}).
This introduces some randomness into the minimization
procedure which might help to overcome local minima but introduces at the same
time
a new free parameter $\tilde N$ which can only be optimized by trials.
For $\tilde N\:=\:N$ the method is called \mbox{\it batch mode updating}.\\
For standard FFBP the temperature $t$ does not affect the performance of
the minimization procedure. As for the entries of the matrices, no
constraints are imposed, any change of $t$ can be compensated by an
overall rescaling of the weights.\\

The least squares fit procedure requires several input parameters whose values
are
to be assumed and tested, i.e.
\begin{itemize}
\item the temperature $(t)$,
\item the initial range of the weights $(\epsilon)$,
\item the weight of the momentum term $(\kappa)$,
\item the step size $(\eta)$, and
\item the number of patterns over which the sum for $E$ extends $(\tilde N)$.
\end{itemize}

\bigskip
The performance of the FFBP is strongly influenced by the choice of
these parameters. For instance the possibility of getting trapped in a
local minimum is enhanced for a wrong choice of $\epsilon$.
Often the surface of $E$ is probed by initializing the matrices of
$f(\vektor{X})$ at different points in the parameter space and by
choosing different
sets of values for the above listed parameters. Each of the networks will
achieve a
different performance on an independent test sample, which has not been
part of the sample used to minimize the cost function. This usually happens if
the
minimization procedure converges to a local minimum.
It can be shown, that an average over the output values
of the different networks improves significantly the overall performance on the
test sample.
This method is referred to as the {\it ensemble method} \cite{ensemble1}.
However, it does not ensure an optimal solution and the results depend on the
number of trials.
Another approach is denoted as \mbox{\it weight decay} and reduces the number
of weights by adding a penalty term to the cost function. This term depends on
the size of the weights and thus gives each weight a tendency to decay to zero.
Thereby it is less likely that the error function exhibits local minima because
it depends on less weights \cite{weight_decay}.\\

The problem of restricting the parameter space to a region which ensures
convergence to the global minimum remains thus to a large extend unsolved.

\section{Modified FFBP model}
\begin{figure}
\epsfxsize=7.0cm
\hspace{5cm}\epsffile{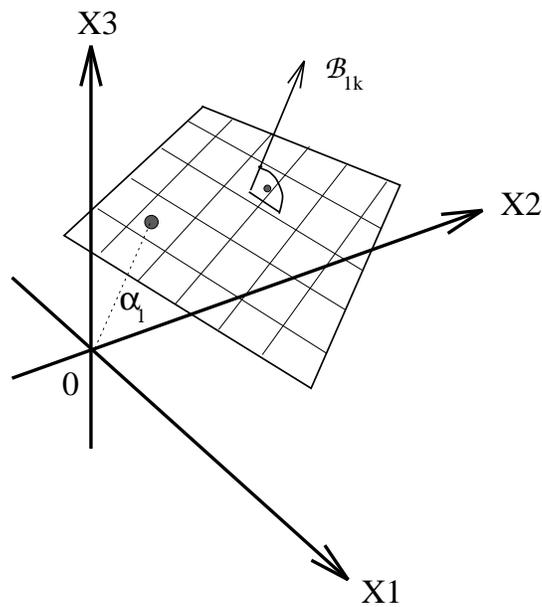}
\caption{Graphical representation of the first transformation of the
neural net formula $(2)$. The matrix operation
$\NM{B}_{jk}\:X_{k}\,-\,\alpha_{j}$ calculates the
Euclidean distances of the input vector \vektor{X} to the $J$
hyperplanes in the space of the input variables \mbox{\it(Hesse's
normal form)}. An example for the first hyperplane $(j\,=\,1)$ in 3
dimensions $(K\,=\,3)$ is depicted. Each row of the matrix
$\NM{B}_{jk}$ denote a normal of a hyperplane in the space of the input
variables. The value $\alpha_{j}$ determines the distance of the $j$'s
hyperplanes to zero.}
\label{plane}
\end{figure}

Let's assume the entries of the input vector \vektor{X} to be of the
order of unity.
The matrices $A_{ij}$ and $B_{jk}$ can be normalized for each row, i.e.
\begin{displaymath}
A_{ij}~\longrightarrow~\NM{A}_{ij}~~~,~~~B_{jk}~\longrightarrow~\NM{B}_{jk}
\end{displaymath}
\begin{equation}
\sum\limits_{j=1}^{J}~\NM{A}_{ij}^{2}~=~1~~~,~~~\forall~i~~~;~~~
\sum\limits_{k=1}^{K}~\NM{B}_{jk}^{2}~=~1~~~,~~~\forall~j~~~~.
\end{equation}
Thereby $\NM{B}_{jk}$ and $\alpha_{j}$ denote
the normal and the Euclidean distance to zero of the $j$'s hyperplane in the
space of the input variables as depicted in figure \ref{plane}. The
hyperplanes are defined by the equations
$\NM{B}_{jk}\:X_{k}\:-\:\alpha_{j}~=~0$.
The same is true for the hyperplanes in the space of the hidden variables,
with the replacement $\NM{B}_{jk}~\rightarrow~\NM{A}_{ij}\:,~
\alpha_{j}~\rightarrow~\beta_{i}$ and $X_{k}~\rightarrow~g_{j}$.
These constraints remove the dependence of the minimization procedure on
$\epsilon$. The weights $\alpha_{j}$ and $\beta_{i}$ are initially set to zero
such
that only the orientation of the hyperplanes vary. Due to this modification the
role of $t$ becomes a major one and rules the overall structure of the error
function. For $t\rightarrow \infty$ the values of $Y_{i}$ are all equal
to 0.5 for any
finite \vektor{X}. The value of $E$ will thus converge to a constant, i.e.
\begin{equation}
\lim_{t\rightarrow\infty}~E~=~\frac{1}{2N}\sum\limits_{n=1}^{N}
                                          \sum\limits_{i=1}^{I}~
\left(~\frac{1}{2}-O_{i}(C_{l},n)~\right)^{2}~=~\frac{I}{8}~~~~.
\end{equation}
In the limit of $t\approx 0$ the sigmoid function $g$ becomes the step
function $\Theta$.
This results in probabilities $Y$ equal to $1$ or $0$, thus in non
overlapping distributions in the input space, i.e. the input distributions
are completely separated.
In terms of the parameter $t$, the error function $E$ acquires a well
defined structure.
The contribution to $E$ of patterns belonging to class $C_{l}$ is determined by
the following expression :
\begin{equation}
E(C_{l})~=~\int d\vektor{X}~P(C_{l})\:P(\vektor{X}|C_{l})~
\frac{1}{2}
\sum\limits_{i=1}^{I}~
\underbrace{
\left(~Y_{i}(\vektor{X}(C_{l}))~-~O_{i}(C_{l})~\right)^{2}
}_{\Lambda(C_{l})}~~~~.
\label{contribution_to_e}
\end{equation}
If an input vector \vektor{X} belongs to class $C_{l}$ the function
$P(\vektor{X}|C_{l})$ denotes the probability distribution of \vektor{X} and
$P(C_{l})$ the probability of class $C_{l}$. Both functions depend upon the
problem
under investigation, thus $\Lambda(C_{l})$ is the only part of equation
(\ref{contribution_to_e}) which changes as a function of the parameter $t$.
To analyse the high $t$-behavior of $E(C_{l})$ one can expand $\Lambda(C_{l})$
in $\frac{1}{t}$.
\begin{eqnarray}
\mbox{Denoting }h_{j} & := & \NM{B}_{jk}\:X_{k}\,-\,\alpha_{j}~~~, \nonumber \\
\lim\limits_{\frac{1}{t}\rightarrow 0}~\Lambda(C_{l}) & = &
\left(~g\left[~\frac{\NM{A}_{ij}\:g(~\frac{h_{j}}{t}~)}{t}~-~\frac{\beta
_{i}}{t}~\right]
{}~-~O_{i}(C_{l})~\right)^{2} \nonumber \\[0.5cm]
 & = & \left(~\frac{1}{2}\left(~1\:+\:\frac{\NM{A}_{ij}}{2\,t}\:+\:
       \frac{\NM{A}_{ij}\,h_{j}}{2\,t^{2}}\:-\:\frac{\beta_{i}}{t}\:+\:
       \NM{O}(\:\frac{1}{t^{3}}\:)~\right)~-~
       O_{i}(C_{l})~\right)^{2} \nonumber \\[0.5cm]
 & = & O_{i}^{2}(C_{l})~-~
       O_{i}(C_{l})~\left(~1~+~\frac{\NM{A}_{ij}}{2\,t}~+~
\frac{\NM{A}_{ij}\,h_{j}}{2\,t^{2}}~-~\frac{\beta_{i}}{t}~\right)~+~
\nonumber\\
 & ~ & \frac{1}{4}~\left(~1+\frac{\NM{A}_{ij}}{t}~+~
       \frac{\NM{A}_{ij}^{2}}{4\,t^{2}}~+~
       \frac{\NM{A}_{ij}\,h_{j}}{t^{2}}~-~\right.\nonumber\\
 & ~ & \mbox{}~~~~~~~\left.2\,\frac{\beta_{i}}{t}~-~
       \frac{\NM{A}_{ij}\,\beta_{i}}{t^{2}}~+~
       \frac{\beta_{i}^{2}}{t^{2}}~\right)~+~\NM{O}(\:\frac{1}{t^{3}}\:)
\label{nn_approximation}
\end{eqnarray}

For small $\frac{1}{t}$, when terms of the order $\NM{O}(\frac{1}{t^{3}})$
or higher can be neglected, the error function becomes a quadratic sum of the
weights with only one minimum. In the next section this will be illustrated
with
an example.\\

Thus, while for high temperatures the error function has a smooth behavior, at
low
temperatures all its structures are present.
This transition is continuous and it is reasonable to assume that the
global minimum of
the error function becomes at high temperatures the only minimum of $E$.
The idea is then to start the minimization at high values of $t$ and
converge to the region of the minimum of $E$ in this regime. The resulting
weights
are expected to be already close to those corresponding to the global minimum.
If
so, further decrease of the temperature should lead to the global minimum
without
the risk of being trapped in a local minimum. Therefore the
summation in equation (\ref{error_function}) must extend over the whole
pattern sample
to determine the position of the global minimum as precisely as possible , i.e.
the
free parameter $\tilde N$ is set to $N$. In the minimization procedure the
temperature is changed in the same way as the weights but the step size for the
temperature is reduced by one order of magnitude. This ensures a faster
convergence behavior for the weights, therefore

\begin{eqnarray}
t^{(\mu+1)}      & = & t^{(\mu)}~-~\frac{\eta}{10}\:\Delta t^{(\mu)} \\
\Delta t^{(\mu)} & = & \frac{\partial E^{(\mu)}}{\partial
t}~+~\kappa\Delta t^{(\mu-1)}~~~~.
\end{eqnarray}

In the low $t$ region $E$ becomes very steep which might lead during
the minimization
to oscillations in $t$. Any step in the wrong direction and the error
function could
yield huge derivatives for $t$. Thus given that the functional
dependence of $E$ on $t$
is of the form $E\:\sim\:\tanh^{2}\frac{1}{t}$, the step size $\eta$ must be
$t$-dependent, i.e.
\begin{equation}
\eta(t)~=~1+~\gamma~-~\tanh^{2}\frac{1}{t}~~~~,
\end{equation}
where the new parameter $\gamma$ is the pedestal value for $\eta(t)$.
A value of $\gamma$ of $0.1$ ensures that only about $10\%$ of the calculated
derivatives will contribute to the change of the weights for $t\:\leq\:0.5$.\\

\subsection{Determination of $t^{(0)}$}
\begin{figure}[t]
\epsfxsize=10cm
\hspace{3cm}\epsffile{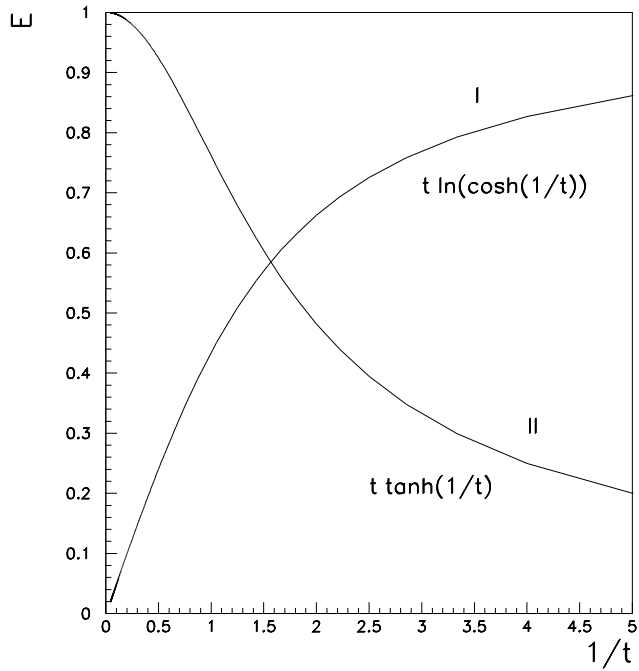}
\caption{The t-dependence of the individual terms contributing to the error
function $E_{ij}(C_{l})$ for a one-dimensional
distribution with a flat conditional probability.}
\label{demo2}
\end{figure}

Usually one aims to separate two distinct distributions. In that case the
neural net formula (\ref{nn_formula}) can be simplified. We set $I\:=\:1$,
change the sigmoid function to $g(...)\:=\:\tanh(...)$ and assign two
output values to the now scalar variable $O(C_{l})~:$
\begin{displaymath}
O(C_{l})~=~\left\{~\begin{array}{l}
                     +1~,~\mbox{if \vektor{X} belongs to class $C_{1}$} \\
                     -1~,~\mbox{if \vektor{X} belongs to class $C_{2}$}
                    \end{array}
            \right.~~~~.
\end{displaymath}
Therefore $Y$ is a scalar and becomes an estimator for the probability function
\begin{equation}
Y~\approx~P(C_{1}|\vektor{X})\:-\:P(C_{2}|\vektor{X})~~~~,
\end{equation}
with $P(C_{l}|\vektor{X})$ being the a posteriori Bayesian probability
that \vektor{X}
belongs to the class $C_{l}$. For the case of two overlapping Gaussian
distributions
$g(...)$ represents the exact solution for the Bayesian probabilities. If we
consider a one-dimensional problem $(K\,=\,1)$ and allow for simplicity just
one cut $(J\,=\,1)$ formula (\ref{nn_formula}) can be reduced to :
\begin{equation}
Y(X(C_{l}))~=~
\tanh\left[~\frac{X(C_{l})\:-\:d}{t}~\right]~~~~,
\label{one_dim_nn_formula}
\end{equation}
with $d$ as the only remaining weight, corresponding to $\alpha_{j}$ in
equation (\ref{nn_formula}). Let us assume a flat conditional probability
distribution for the different classes, i.e.
\begin{equation}
P(X|C_{l})_{ij}   ~=~\left\{~\begin{array}{lll}

\frac{1}{X_{j}-X_{i}}&,&\mbox{if}~X_{i}\,\leq\,X\,\leq\,X_{j} \\
                     0                    &,&\mbox{elsewhere}~~~~~~~~~~~~.
                    \end{array}
            \right.
\label{flat_prob_dist}
\end{equation}
The contribution of $P(X|C_{l})_{ij}$ to the error function $E$ can be
evaluated analytically :
\newpage
\begin{eqnarray}
E_{ij}(C_{l})  & = & \frac{1}{2}\:\frac{P(C_{l})}{X_{j}-X_{i}}~
                     \left[~
                         O^{2}(C_{l})\,(X_{j}-d)\,-\,
2\,t\,O(C_{l})\,\ln\cosh\frac{X_{j}-d}{t}\,+\right.\nonumber \\
               & ~ &     ~~~~~~~~~~~~~~~~~~~~
                         \left. (X_{j}-d\:-\:t\,\tanh\frac{X_{j}-d}{t})~
                     \right]~- \nonumber \\
               & ~ & \frac{1}{2}\:\frac{P(C_{l})}{X_{j}-X_{i}}~
                     \left[~
                         O^{2}(C_{l})\,(X_{i}-d)\,-\,
2\,t\,O(C_{l})\,\ln\cosh\frac{X_{i}-d}{t}\,+\right.\nonumber \\
               & ~ &       ~~~~~~~~~~~~~~~~~~~~
                         \left. (X_{i}-d\:-\:t\,\tanh\frac{X_{i}-d}{t})~
                     \right]~~~~.
\label{one_dim_e}
\end{eqnarray}
For high values of $t$ the non-linear functions in equation (\ref{one_dim_e})
are expanded in powers of $(\frac{X-d}{t})$. Neglecting terms of the order
of $\NM{O}(\:(\frac{X-d}{t})^{5}\:)$ and higher will leave a $d^{4}$
dependence in $E_{ij}(C_{l})$ which might lead to a local minimum. It's
partial derivative relative to $d$ turns out to be only quadratic :
\begin{eqnarray}
\partial_{d}\,E_{ij}(C_{l}) & = & P(C_{l})~\left(~
d^{2}\,\left(-\frac{O(C_{l})}{t^{3}}\right)~+~
d\,(\ldots)~+~(\ldots)
{}~\right) \\
\partial_{d}\,E & = & \partial_{d}\,E_{ij}(C_{1})~+~
                      \partial_{d}\,E_{kl}(C_{2})
\end{eqnarray}
This already proves that $E$ has only one maximum and one minimum. The
quadratic
dependence of the partial derivative $\partial_{d}\,E$ on the weight
$d$ vanishes for $P(C_{l})\,=\,0.5$ due to the definition of $O(C_{l})$.
With this additional requirement the error function $E$ depends only
quadratically
on the weight $d$.
Thereby the optimal a priori probability for the minimization
procedure is determined to be 0.5. As $Y$ will be an estimator for an
a posteriori probability it is possible afterwards to reweigh the result
to any a priori probability under investigation.
The $t$-dependence of $E_{ij}(C_{l})$ is determined by a sum of 2 terms
\begin{equation}
E_{ij}(C_{l})~\sim~\underbrace{t\,\ln\cosh\frac{1}{t}}_{I}~+~
\underbrace{t\,\tanh\frac{1}{t}}_{II}~~~~,
\end{equation}
which are depicted in figure \ref{demo2}. The two terms exhibit a
different behavior for $t\rightarrow 0$ and $t\rightarrow \infty$.
The first term $(I)$ dominates in the
low $t$ region and therefore determines the structure
of the error function $E$ in this temperature range.
For high values of $t$ the second term $(II)$
dominates over the first one. Both terms have almost equal weight for
$1/t\:=\:1.5$.
Since the proposed method of minimizing the error function requires that
initially $t^{(0)}$ be chosen such that $E$ exhibits only one minimum, it
implies $1/t^{(0)}\:<\:1.5$.
At $t\:=\:5$ the second term $(II)$ has a 10-times bigger weight than the first
term $(I)$ which should satisfy the requirements necessary for the
approximation
done in equation (\ref{nn_approximation}). With the assumption that the values
of
the input vector \vektor{X} are of order one, the general prescription for
the initial value of the temperature $t$ is thus
\begin{equation}
t^{(0)}~\geq~5~~~~.
\label{initial_temp}
\end{equation}

\begin{figure}[b]
\epsfxsize=10cm
\hspace{3cm}\epsffile{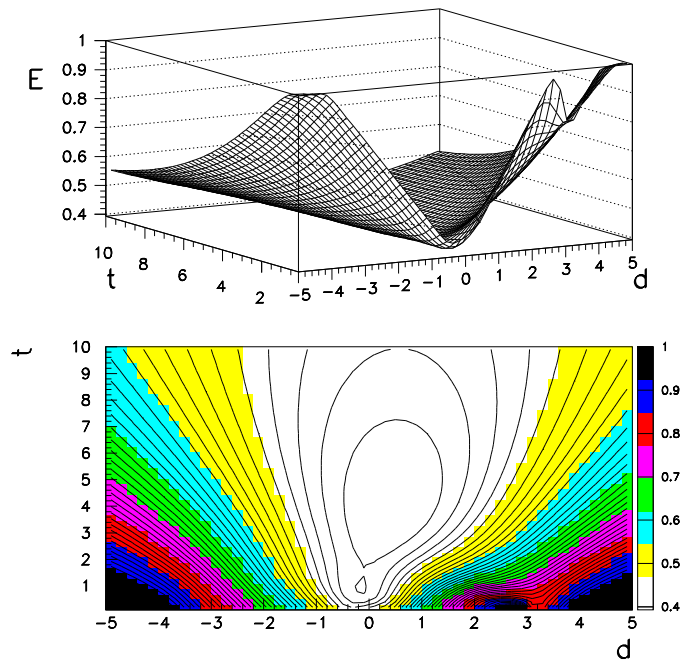}
\caption{Structure of the error function $E$ for the one-dimensional example
(described in the text) as a function of the weight $d$ and the
temperature $t$. The
global minimum of this specific error function is placed at $d\:\approx\:0.5$
and
$t\:\approx\:4.5$. Local minima of $E$ lie below $t\:\approx\:1.3$.}
\label{demo1}
\end{figure}

This will be illustrated by a numerical example. Suppose one aims to separate
the
two one-dimensional overlapping distributions with flat conditional
probabilities
as defined in equation (\ref{flat_prob_dist}) and assumes that :
\begin{eqnarray*}
P(X|C_{1}) & = & 0.7\:P(X|C_{1})_{12}~+~0.3\:P(X|C_{1})_{34} \\
P(X|C_{2}) & = & 1.0\:P(X|C_{2})_{56}
\end{eqnarray*}
\begin{displaymath}
\begin{array}{lll}
X_{1}~=~-4.0 & X_{2}~=~0.5 & X_{3}~=~3.0 \\
X_{4}~=~3.1  & X_{5}~=~-0.5 & X_{6}~=~4.0
\end{array}
\end{displaymath}
With $P(C_{1})\,=\,P(C_{2})\,=\,0.5$ one gets a surface of the error function
$E$ as depicted in figure \ref{demo1}. If the minimization procedure were to
start
at any value of $d$ and $t^{(0)}~\geq~5$ it would converge to the global
minimum of $E$ without the possibility of getting trapped in a local minimum.
In the case of non-overlapping distributions the temperature $t$ will
converge to zero to model probabilities equal to one. If the distributions
overlap to $100\,\%$ the temperature will converge to infinity. Thereby the
final value of $t\,=\,t^{(\infty)}$ becomes a measure of the overlap of the two
distinct distributions, i.e.
\begin{equation}
\mbox{\rm Overlap} ~\approx~\frac{t^{(\infty)}}{1~+~t^{(\infty)}}~~~~.
\end{equation}

Thus to summarize, the proposed modified neural network differs from networks
using
standard backpropagation as follows :

\begin{center}
\small
\begin{tabular}{|c|l|l|l|}
\hline
Parameter        & Name          & New model    & Standard FFBP   \\
\hline
                 &               &                       &     \\[-5pt]
$t^{(0)}$        & The initial value
                                 & $t^{0}\,\geq\,5~.$ The temperature is {\bf
not}
                                                & Not well defined. \\
                 & of the temperature
                                 & constant and changes for
                                                & Usually $t$ is not changed \\
                 &               & each iteration step.
                                                & throughout the \\
                 &               &              & minimization procedure\\
                 &               &              & and thus set to $t\:=\:1$. \\
\hline
$\epsilon$       & The absolute range
                                      & Cancelled, as the weights \NM{A} and
\NM{B}
                                      & Not well defined.  \\
                 & of the initial& are row-wise normalized to one,
                                      &  Usually $\epsilon\,\leq\,0.01$ \\
                 & weights       & and $\alpha$ and $\beta$ are
initially set to zero
                                      &  \\
\hline
$\kappa$         & The weight of the
                                 & Not well defined      & Not well defined \\
                 & momentum term &                       & \\
\hline
$\eta$           & The step size
                                 & $\eta(t)\:=\:1.0\,+
                                   \,\gamma\,-\,\tanh^{2}\frac{1}{t}$
                                                         & Not well defined. \\
                 &               &                      & Usually
$\eta\,\leq\,0.001$\\
\hline
$\gamma$         & The pedestal for $\eta$
                                 & Not well defined. Usually set to $0.1$
                                                         & Does not exist in \\
                 &               &                       & this model \\
\hline
                 &               &                       &     \\[-5pt]
$\tilde N$       & The \# of patterns
                                 & $\tilde N \,=\,N$     & Not well
defined. \\                         & to sum over in
                   $E$           &                       & Usually set to
                                                           $\tilde
N\,\approx\,10$ \\
\hline
$P(C_{l})$       & The a priori  & For the case of two classes
                                                         & Not well defined \\
                 & probability   & determined to $P(C_{l})\,=\,0.5$,
                                                         &     \\
                 & of class $C_{l}$
                                 & otherwise not well defined
                                                         &     \\
\hline
\end{tabular}
\end{center}

\section{Conclusions}
A novel method to minimize the quadratic cost function of a neural network
with error backpropagation has been presented. The essential modification
is the row-wise normalization of the matrices \NM{A} and \NM{B}
which represent part of the weights of the neural network. Thus, the entries
of each row of \NM{A} and \NM{B} acquire the meaning of
normals which define the orientation of hyperplanes. Due to
the normalization,
the error function $E$ obtains a well defined structure as a function of
the free parameter $t$, denoted as the temperature. It has been proven
that for high values of $t$, when terms of the order $\NM{O}(\frac{1}{t^{3}})$
or higher can be neglected, the cost function $E$
always exhibits a quadratic dependence on the
weights and thus only one minimum. For low temperatures
all structures of $E$ are apparent and local minima might exist.
This transition is continuous and it is natural to assume, that the single
minimum of the cost function at high values of $t$ leads to the global
minimum of $E$ at low temperatures. However, there is no rigorous proof as yet,
that this should be the case.

The minimization procedure starts at high temperatures and converges first to
the single minimum of the error function in this range of $t$.
Further decrease of $t$ should lead to the global minimum
without the risk to converge to a local minimum. Similar to the weights,
the temperature becomes a parameter whose value is determined by the
minimization procedure. Assuming the entries of the input vector \vektor{X}
to be of the order of one, the initial value of $t$ should be
$t^{(0)}\:\geq\:5$
as derived from an one-dimensional example.
Multi-dimensional problems are nested superpositions of the one-dimensional
example, therefore this range for $t^{(0)}$ should still ensure the quadratic
dependence of the cost function on the weights.
Several free parameters of the standard minimization procedure for FFBP,
whose values are to be assumed and tested, are constrained. Thus, without any
fine
tuning, the new model is applicable to any classification problem.\\

The new method described in this paper has been successfully applied to the
problem of electron identification~\cite{my2} in the ZEUS
ldetector~\cite{ZEUSdet} at HERA. At HERA electrons of 30 GeV collide with
protons of 820 GeV.  High energetic particles and jets of particles, mainly
hadronic particles, emerge from the interaction point and deposit energy
in the spatially segmented uranium-scintillator calorimeter (CAL) of the ZEUS
detector. A certain fraction of the interactions between electrons and
protons are characterized by the presence in the final state of the electron
scattered under a large angle, which is thus in the geometrical acceptance
of the CAL. The aim is to select this type of events which are believed to
originate from the scattering of the electron on a point like constituent of
the proton. The showering properties of electrons and hadrons in an absorber
material are different and it is possible to identify the particle type by
the pattern of the energy deposits in the CAL. The longitudinal and
transversal segmentation of the CAL provides 54 values reflecting the
spatial shape of the shower. The shape depends also on the angle of incidence
of the showering particle. After including this angle the input
patterns are 55-dimensional. Using the new method, a neural network has been
trained on patterns originating from electrons and hadrons. In comparison
with a classical approach, the neural network separates the distinct
distributions better, giving a typical increase of about 10\% in efficiency
and purity. A principle component analysis has shown that this
improvement is achieved through the use of all the 55 variables.

\vspace{2cm}

\noindent {\Large\bf Acknowledgments}

This work has been pursued in the framework of the ZEUS collaboration and
I would like to acknowledge the motivating interest of my colleagues. In
particular I would like to thank Prof. Halina Abramowicz for her strong support
and the many fruitful discussions. I am indebted to Prof. David Horn for his
many
helpful suggestions. I would like to thank Prof. Klaus Wick for his
encouragement.

\section{Appendix}
This new minimization procedure has been implemented in a {\tt FORTRAN} program
called {\tt dEXTRa}.
The program is controlled by a configuration file named {\tt dextra.cnf} where
all parameters, paths and options are set. It must be located in the directory
from which {\tt dEXTRa} is started. The value of the error function $E$
for $t\,=\,0$
for both the training and an independent test pattern sample can be
calculated during
the minimization procedure. This method is denoted as \mbox{\it
cross-validation}
and serves to check for over-fitting, i.e. when the network starts to pick up
fluctuations from the training sample \cite{cross_validation}. After the
training
procedure the final set of matrices and parameters can be written to a file
which
afterwards can be read in again for application to new pattern samples.\\

The program is available on request from the author.
For further questions please contact \mbox{\tt sinkus$@$zow.desy.de}.

\bibliographystyle{plain}
\bibliography{neural}

\clearpage

\clearpage

\end{document}